\documentclass[useAMS,usenatbib]{mn2e}

\usepackage{graphicx}
\usepackage{epsfig}

\newcommand{\GA}{\mbox{\raisebox{-0.6ex}{$\stackrel{\textstyle>}{\sim}$}}}
\begin{document}
\title{General Relativistic Considerations of the Field Shedding Model of Fast Radio Bursts}
\author[Brian Punsly and Donato Bini] {Brian Punsly and Donato Bini\\ 1415 Granvia Altamira, Palos Verdes Estates CA, USA
90274 and ICRANet, Piazza della Repubblica 10 Pescara 65100, Italy\\
\\Istituto per le Applicazioni del Calcolo \lq\lq M. Picone,"  C.N.R., I-00185 Rome,
Italy\\ and International Center for Relativistic Astrophysics,
I.C.R.A., University of Rome La Sapienza, I-00185 Roma, Italy\\
\\E-mail: brian.punsly1@verizon.net}
\maketitle \label{firstpage}
\begin{abstract} Popular models of fast radio bursts (FRBs) involve the gravitational collapse of neutron
star progenitors to black holes. It has been proposed that the
shedding of the strong neutron star magnetic field ($B$) during the
collapse is the power source for the radio emission. Previously,
these models have utilized the simplicity of the Schwarzschild
metric which has the restriction that the magnetic flux is magnetic
``hair" that must be shed before final collapse. But, neutron stars
have angular momentum and charge and a fully relativistic Kerr
Newman solution exists in which $B$ has its source inside of the
event horizon. In this letter, we consider the magnetic flux to be
shed as a consequence of the electric discharge of a metastable
collapsed state of a Kerr Newman black hole. It has also been argued
that the shedding model will not operate due to pair creation. By
considering the pulsar death line, we find that for a neutron star
with $B = 10^{11} - 10^{13}$ G and a long rotation period, $>1$ s
this is not a concern. We also discuss the observational evidence
supporting the plausibility of magnetic flux shedding models of FRBs
that are spawned from rapidly rotating progenitors.
\end{abstract}

\section{Introduction}

Fast radio bursts (FRBs) are a new kind of astrophysical transient
that is of unknown origin. First discovered in \citet{lor07}, it was
later found that the distribution lies primarily well above the
Galactic plane. Combined with a large dispersion measure indicates a
celestial origin and large distances corresponding to redshifts,
$0.5 <z <1$ \citep{tho13}. The radio bursts durations are on the
order of milliseconds with a radio luminosity of $10^{38} - 10^{40}$
ergs \citep{zha14}. The space density is large, $ 10^{-3}\,
\mathrm{gal}^{-1} \, \mathrm{yr}^{-1}$ compared to gamma ray bursts
(GRBs), a factor of 1000 larger \citep{zha14}. Light travel time
arguments based on 1 msec imply that the emission region is very
compact, less than 20 times a neutron star (NS) radius or 30 times
the diameter of a black hole (BH) of a few solar masses. This
inspired the ``blitzar" model of FRBs that is based on NS collapse
to a BH, \citet{fal14}, that was later adapted to the work of
\citet{zha14}. The models begin with a supramassive NS that is
marginally supported by centrifugal force. During the collapse, the
magnetic ``hair" is released, the so-called ``no hair theorem." This
is the magnetic energy source for the FRB in the model. It has been
largely forgotten, but the model of shedding magnetic hair was
originally proposed to explain GRBs \citep{han97}.
\par This letter is concerned with the two fundamental
physical points of concern for the model. Firstly, previous FRB
models have considered a Schwarzschild metric. With this
restriction, the magnetic field cannot have its source within the
black hole and the field is considered magnetic hair that needs to
shed during the collapse. In reality, there is angular momentum and
charge and the Kerr-Newmann solution (a charged rotating black hole,
KNBH hereafter) allows for a magnetic field that has its source
inside the event horizon. The neutron star is charged and rotating,
so is the black hole. The second issue is the notion that the
magnetic flux might not be released since pair creation in the
magnetosphere might self-sustain the field even after collapse
\citep{lyu11}. We find an interpretation involving a progenitor NS
beyond the ``pulsar death line" and electric discharge of a
meta-stable KNBH intermediate state that allows the magnetic field
shedding model of FRBs to proceed regardless of these concerns.
Heavy dead pulsars may require centrifugal force to be supported
against gravitational collapse. We argue that that the torque down
of dead pulsars could initiate collapse with a space density
consistent with the FRB space density. We also comment on whether a
supramssive rapidly rotating NS can collapse to form a meta-stable
KNBH.

\section{The Kerr-Newman Field Energy}
The magnetic field energy responsible for the FRB was calculated
outside a NS in the Newtonian approximation \citep{zha14}. In this
section, we calculate the magnetic field energy and the electric
field energy in the Kerr-Newman spacetime that represents a rotating
charged black hole with a net magnetic flux through each hemisphere.
The axisymmetric, time stationary spacetime metric is uniquely
determined by three quantities, $M$, $a$ and $Q$, the mass, angular
momentum per unit mass, and the charge of the hole respectively. In
Boyer--Lindquist coordinates the metric, $g_{\mu\nu}$, is given by
the line element in geometrized units

\begin{eqnarray}
&&\mathrm{d}s^{2} \equiv g_{\mu\nu}\, \mathrm{d}x^{\mu}\mathrm{d}
x^{\nu}= -\left (1-\frac{2Mr - Q^{2}}{\Sigma}\right)
\mathrm{d}t^{2}+\Sigma\mathrm{d}\theta^{2} \nonumber \\
&& +\left(\frac{\Sigma}{\Delta}\right)\mathrm{d}r^{2} -\frac{(4Mr -
2Q^{2})a}{\Sigma}\sin^{2}\theta\, \mathrm{d}\phi \,
\mathrm{d}t  \nonumber \\
&& +\left [(r^{2}+a^{2})+\frac{(2Mr - Q^{2})a^{2}}{\Sigma}\sin^{2}
\theta\right ] \sin^{2} \theta \, \mathrm{d}\phi^{2} \; ,
\end{eqnarray}
where $\Sigma = r^{2}+a^{2}\cos^{2}\theta$ and
\begin{equation}
\Delta = r^{2}-2Mr+a^{2}+Q^{2} \equiv
\left(r-r_{{+}})(r-r_{{-}}\right ).
\end{equation}

There are two event horizons given by the roots of the equation
$\Delta=0$. The outer horizon at $r_{{+}}$ is of physical interest
\begin{equation}
r_{{+}}=M+\sqrt{M^{2}-Q^{2}-a^{2}} \; .
\end{equation}
We choose to simplify our calculations by computing quantities in a
hypersurface orthogonal, orthonormal frame. A natural orthonormal
frame associated with Zero Angular Momentum Observers (ZAMO) (who
are also locally non-rotating) can be used to express, locally, the
electromagnetic field in terms of electric and magnetic
(observer-dependent) fields. Being hypersurface orthogonal (i.e.,
vorticity-free), the ZAMO frame gives an unambiguous definition of
the field that is integrable \citep{pun08}. The basis covectors are

\begin{eqnarray}
&&\omega^{\hat 0}=1/\sqrt{-g^{tt}}dt\; ,\; \omega^{\hat
r}=\sqrt{g_{rr}}dr\; , \nonumber \\ && \omega^{\hat
\theta}=\sqrt{g_{\theta\theta}}d\theta\;, \;\omega^{\hat
\phi}=\sqrt{g_{\phi\phi}}d\phi \;.
\end{eqnarray}
Due to the long expressions to follow, it is worthwhile to
abbreviate the notation for $(\sin\theta, \cos\theta)$ as $(s, c)$,
i.e. $\Sigma=r^2+a^2c^2$.

\par In the ZAMO basis, the Kerr-Newman electric and magnetic fields
are

\begin{eqnarray}
E(n)_{\hat r} &=& +\frac{Q(r^2+a^2)(r^2-a^2c^2)}{\Sigma^2\sqrt{(r^2+a^2)^2-a^2s^2\Delta}}\nonumber \\
E(n)_{\hat \theta} &=&- \frac{2Qra^2
sc\sqrt{\Delta}}{\Sigma^2\sqrt{(r^2+a^2)^2-a^2s^2\Delta}}
\nonumber \\
B(n)_{\hat r} &=&  -\frac{2arcQ(r^2+a^2)}{\Sigma^2\sqrt{(r^2+a^2)^2-a^2s^2\Delta}}\nonumber \\
B(n)_{\hat \theta} &=&
-\frac{aQs\sqrt{\Delta}(r^2-a^2c^2)}{\Sigma^2\sqrt{(r^2+a^2)^2-a^2s^2\Delta}}\,.
\end{eqnarray}
The associated electromagnetic invariants are
\begin{eqnarray}
{\mathcal I}_1&=& [E(n)]^2-[B(n)]^2=Q^2 \frac{(r^2-2 r ac-a^2c^2) (r^2+2 r ac-a^2c^2)}{\Sigma^4}\nonumber\\
{\mathcal I}_2&=& E(n)\cdot B(n)= -Q^2 \frac{2arc
(r^2-a^2c^2)}{\Sigma^4}\;.
\end{eqnarray}
Similarly, the ZAMO-relative energy density is
\begin{eqnarray}
{\mathcal E}(n)
&=& \frac{1}{8\pi} [E(n)^2+B(n)^2]\nonumber \\
&=& \frac{1}{8\pi}
\frac{Q^2}{\Sigma^2}\left(\frac{(r^2+a^2)^2+a^2s^2\Delta}{(r^2+a^2)^2-a^2s^2\Delta}\right)\,.
\end{eqnarray}
The 3-volume element, $dV$, and the 2-volume element, $dA$, at fixed
r and for $\theta\in [0,\pi]$ coordinate are needed in the following
\begin{eqnarray}
&& dV = \sqrt{g_{rr}g_{\theta\theta}g_{\phi\phi}}\, dr\, d\theta\,
d\phi\nonumber \\
&& = \sin\theta \sqrt{\Sigma[(r^2+a^2)^2-a^2s^2\Delta]}/\sqrt{\Delta}\, dr\, d\theta\, d\phi\,,\nonumber\\
&&dA = \sqrt{g_{\theta\theta}g_{\phi\phi}}\, d\theta\, d\phi
\nonumber\\ && =\sin\theta \sqrt{(r^2+a^2)^2-a^2s^2\Delta}\,
d\theta\, d\phi\,.
\end{eqnarray}

In order to make contact with the progenitor NS magnetic field, we
compute the magnetic flux through the northern hemisphere of a
sphere of radius r using Equations (5) and (8)

\begin{eqnarray}
\Phi (B) &=&\int_0^{2\pi}\int_{0}^{\pi/2}
 B(n)_{\hat r}\, dA =
\frac{2\pi aQ}{r}\,.
\end{eqnarray}
If we introduce the redshift between the ZAMO frames and the
stationary observers at asymptotic infinity, the lapse function
($\alpha = \Delta^{1/2}s/(g_{\phi\phi})^{1/2}$), the integrated
field energy is
\begin{equation}
I=\int \alpha {\mathcal E}(n) dV = 2\pi\int_{r_+}^\infty dr
\int_0^\pi d \theta {\mathcal E}(n) \Sigma \sin\theta\,.
\end{equation}
It results after the integration over the $\theta$ coordinate to
\begin{equation}
I=I_1+I_2\,,
\end{equation}
where
\begin{eqnarray}
I_1 = -\frac{Q^2}{a^2}\int_{r_+}^\infty dr \,
\frac{(r^2+a^2)}{(2Mr-Q^2)}\,
\Psi\left(\sqrt{\frac{a^2\Delta}{(r^2+a^2)^2-a^2\Delta }}
\right )\nonumber \\
I_2 = \frac{Q^2}{2a^2} \int_{r_+}^\infty dr \,
\left[\frac{(2(r^2+a^2)-(2Mr-Q^2))}{ (2Mr-Q^2)}\right]\,
\Psi\left(\frac{a}{r}\right)
\end{eqnarray}
with $\Psi(x) \equiv x\arctan (x)$. $I_1$ is the stored energy in
the electric field and $I_2$ is the stored energy in the magnetic
field. These integrals can be approximated in the limit $(a/M)^{2}
\ll 1$, $(Q/M)^{2} \ll 1$ and   $\mid Qa\mid/(M)^{2} \ll 1$, by
$I_{\rm approx} = {\mathcal E}({\rm magn}) + {\mathcal E}({\rm
elec})$,
\begin{eqnarray}
{\mathcal E}({\rm elec}) \approx {\frac {{Q}^{2}}{4M}}+{\frac {{Q}^{4}}{16{M}^{3}}}+{\frac {{a}^{2}{Q}^{2}}{32{M}^{3}}}+\ldots\nonumber \\
{\mathcal E}({\rm magn}) \approx {\frac
{{a}^{2}{Q}^{2}}{32{M}^{3}}}+\ldots \,.
\end{eqnarray}

\section{Collapse of the Neutron Star}

In this section, we discuss the assumptions used in our study of
neutron star collapse in the field shedding model.

\begin{enumerate}
\item The initial B field is approximately dipolar at the NS
surface.
\item The flux stays frozen into the superconducting fluid of
subatomic particles as it collapses through the horizon.
\item The mass and angular momentum are approximately conserved
during the collapse.
\end{enumerate}

One problem with analyzing these assumptions with numerical
simulations is that the fundamental physics is not well understood.
The final result is driven by assumptions of the numerical code and
imposed expediences used to avoid complicated physics. Thus, there
are a variety of disparate outcomes. As mentioned in the
introduction, the flux shedding model of FRBs and GRBs in
\citet{han97,fal14,zha14} take place in a Schwarzschild background
and the authors identify the magnetic flux as ``hair". However, if
the progenitor NS has a magnetic flux, $\Phi_{NS}$, through the
northern hemisphere, by Equation (9) and assumptions (2) and (3),
above, the NS with angular momentum per unit mass, $a_{NS}$, and
mass, $M_{NS}$, can, in principle, collapse to a KNBH defined by
\begin{equation}
Qa=\frac{\Phi_{NS}}{2\pi r_{+}}\,,\quad \: a = a_{NS}\,,\quad\: M=M_{NS}\;,
\end{equation}
without violating any principles of relativity. Thus, there is no
``hair" to shed.
\par It was proposed in \citet{lyu11}, that pair
creation similar to that in the standard pulsar model will continue
to seed the magnetosphere during collapse and source the poloidal
field eternally after the black hole forms. Thereby transferring the
field source form the NS matter to an external magnetospheric pair
plasma. The steady state is a ``boot-strapping" model in which the
pairs create the electromagnetic field which in turn create the
pairs. However, force-free simulations of the collapse of a rotating
NS found that the magnetic flux is shed during collapse due to
reconnection \citep{leh12}. Also, using a variety assumptions both
\citet{han97} and \citet{dio13} showed an electromagnetic pulse that
accompanies the shedding of the magnetic flux before the current
source penetrates the event horizon. However, the background metric
is not altered by the charge and angular momentum of the collapsing
matter, so it does not represent the allowed KN solution in which
magnetic flux shedding need not occur.
\begin{figure}
\begin{center}
\includegraphics[width=85 mm, angle= 0]{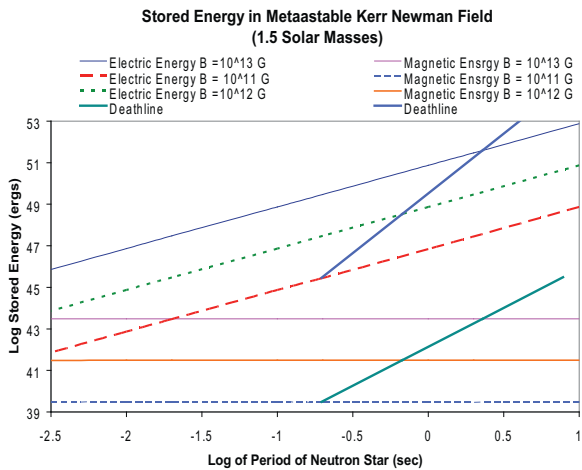}
\caption{The stored electric and magnetic energy of the metastable
Kerr-Newman black hole as a function of the progenitor neutron star,
$B_{NS}$ and $P$. The pulsar death-lines are superimposed based on a
dipolar NS magnetic field. The stored energy increases for
supramassive progenitors with larger radii.}
\end{center}
\end{figure}
\par There is simply not enough known about the physics to resolve
this issue at present. However, we seek a subset of solution space
that circumvents the objection posed by \citet{lyu11}. We consider
the notion of a death-line for pulsars \citep{che93}. The idea is
that the energy available to produce electron-positron pairs is the
voltage drop across magnetic field lines, $\Delta V$, which scales
like $\Delta V \sim B/P_{NS}$ as a consequence of the frozen-in
condition in a rotating magnetosphere (where $P_{NS}$ is the pulsar
rotational period). So for any pair creation mechanism, as $B$
decreases or $P_{NS}$ increases, there will be insufficient energy
to initiate the pair creation that seeds the magnetosphere with
plasma. There will be no observable pulsar. Particle-in-cell
simulations of rotating field aligned dead pulsars indicate that the
magnetic field is not in the force-free configuration, as assumed in
the simulations of \citet{leh12}, but is in a dipolar configuration
\citep{che14}. This agrees with our first assumption in this
section. It is proposed that even though $B/P_{NS}$ increases during
the collapse, by assumptions (2) and (3), the fact that there is no
pre-existing plasma-filled magnetosphere will not allow the change
in $\Delta V$ to effectuate plasma filling of the magnetosphere
before the subatomic fluid passes out of causal contact. Thus, a
KNBH might be able to form, but likely for only a brief instant as
we discuss below.
\section{The Kerr-Newman Metastable State}
As the superconducting fluid pulls its magnetic flux inward during
the collapse, a charge separation occurs. The faster rotation and
larger B implied by assumptions (2) and (3) create a cross field
potential, $\Delta V$, the same effect described above in the pair
creation discussion above. The difference is that within the dense
subatomic material there are numerous collisions that allows charge
to separate instantaneously across the B field in order to keep the
plasma frozen-in (the unipolar inductor affect) \citep{pun08}. This
process proceeds promptly regardless of the tenuous pair plasma in
the magnetosphere (which is unable to cross the field lines). Thus,
the dense subatomic fluid is likely to pass within the event horizon
with its magnetic flux frozen-in, unimpeded by the physics of the
tenuous magnetospheric plasma. The increase of the intrinsic charge
density of the collapsed matter implies that charge is ejected
outward by the unipolar inductor (i.e., the unipolar driven
current). The net result is that the local physical process will
increase the magnitude of the total charge of the collapsed object.
The same physics causes a neutron star to have a net charge in a
pulsar \citep{rud75}. The charge that is added to the subatomic
fluid as a consequence of the charge separation driven by unipolar
induction during the collapse increases the charge from the NS value
to that of the KNBH in Equation (14), $Q_{BH} \sim
[(MR_{NS})/(a^{2})]Q_{NS}$.
\par The increase in the charge of the subatomic fluid described above results in a major difference between the NS electromagnetic field and the
KNBH electromagnetic field. The NS electromagnetic field is magnetic
($E^{2} - B^{2} <0$), meaning that there exist local physical
observers that can see the field as purely magnetic, but there is no
observer that can see a purely electric field. Conversely, by
Equation (6), the KNBH electromagnetic field is electric everywhere
\citep{pun99}. Since there is an unscreened mono-polar electric
field, the KNBH is subject to electric discharge. It is the electric
discharge that sheds the magnetic field by Equation (14). Note that
in the case of exactly zero angular momentum, the Schwarzschild
black hole, there is no solution for a magnetic flux source inside
of the event horizon that is allowed in general relativity and the
no hair theorem is the appropriate interpretation of the flux
shedding. In spite of the argument in the previous section, with our
current level of understanding of the relevant microphysics, one can
not say if it is possible for the discharge to take place before the
black hole forms. However, based on the discussion of unipolar
induction above, we consider this unlikely.
\par In this letter, we consider a prompt discharge of the KNBH,
i.e., it is a metastable configuration that occurs as part of the
collapse. We plot the stored electric and magnetic energy in Figure
1 using Equation (13). We assume a low centrifugal force
configuration of the progenitor NS with collapse to a BH occurring
once the Chandrasekhar limit is exceeded by unspecified interactive
processes with the surrounding medium, $M = 1.5M_{\odot}$. A
supramassive NS progenitor as in \citet{fal14,zha14} would have a
larger mass $M \GA 2.5M_{\odot}$ and a lerger progenitor radius. The
stored energy is plotted as a function of the NS magnetic field
strength at the pole, $B_{NS}$, in the dipolar approximation. The
assumed radius of the NS is 12 km. Note that the magnetic energy is
independent of the rotational period for small $a/M$. The electric
energy increases for smaller $a$, if $B_{NS}$ at the neutron star
pole is held fixed, as a consequence of Equation (13). The relevant
death-line from \citet{che93} is the one for a dipolar magnetic
field that is assumed here. They noted that other magnetic field
configurations can move the death-line to the right, however that
would be inconsistent with the assumptions of the previous
calculations in this letter. For slow rotating NS progenitors there
is an allowed collapsed state KNBH solution for parameters to the
right of the death-line where pair production is suppressed. The
power of the FRBs can be explained for $B_{NS}\GA 10^{12}$ G and
periods longer than 1 second. The rapidly rotating (such as as
supramassive) progenitors require a large centrifugal force to
support them and lie to the left of the death-line. It is also
evident that the stored electric energy exceeds the magnetic energy.
We estimate ${\mathcal E}(\rm{magn})$ and ${\mathcal E}(\rm{elec})$
in Equation (13) in terms of the progenitor NS values in order to
make contact with Equation (1) of \citet{zha14},
\begin{eqnarray}
&&{\mathcal E}({\rm magn}) \approx
\frac{3}{16}\frac{R_{NS}}{M}\left[\frac{1}{6} B_{NS}^{2}
R_{NS}^{3}\right] \approx \frac{1}{8}\left(\frac{a}{M}
\right)^{2}{\mathcal E}({\rm elec})
\nonumber \\
&&{\mathcal E}({\rm elec})\approx \left(\frac{M}{a}
\right)^{2}\frac{3R_{NS}}{2M}\left[\frac{1}{6} B_{NS}^{2}
R_{NS}^{3}\right]\;.
\end{eqnarray}
The magnetic field energy computed in \citet{zha14} is captured in
the last term in square brackets.
\par The logical question is whether this stored energy can be
extracted. Ostensibly, since the energy is electromagnetic it can be
radiated near the speed of light and can avoid being swallowed by
the black hole. We explore if there is a fundamental general
relativistic reason that the energy cannot be radiated by
considering the irreducible mass. The mass of the black hole
decomposes into its rest mass, reducible mass and irreducible mass
\citep{chr71}:
\begin{eqnarray}
&& M^{2}=\left ( M_{\rm irr}+\frac{Q^{2}}{4M_{\rm irr}}\right )^{2}+\left
(\frac{Ma}{2M_{\rm irr}}\right )^{2} \; .
\end{eqnarray}
The irreducible mass cannot be extracted from the black hole. The
remainder, the reducible mass, , $M_{\rm red}=M-M_{\rm irr}$, can be
extracted in principle. From Equations (13) and (16), in the limit
of small $Q/M$ and $a/M$,

\begin{equation}
M_{\rm red}={\mathcal E}({\rm elec})+ {\mathcal E}({\rm magn})
+\frac18 \,{\frac {{a}^{2}}{M}}+{\frac {5}{128}}\,{\frac
{{a}^{4}}{{M}^{3}}} +\frac{1}{32}\,{\frac {{Q}^{2}{a}^{2}}{{M}^{3}}}+...
\;,.
\end{equation}
The electromagnetic stored energy is extractable. The last three
terms in Equation (17) appear to represent the rotational energy.
The first two of these are the rotational energy of a Kerr black
hole. There is an extra term, $\frac{1}{32}\,{\frac
{{Q}^{2}{a}^{2}}{{M}^{3}}}$, that is not present in the Kerr
geometry and is not part of the stored field energy integral in
Equation (13). It appears to a relativistic correction to the first
term of the rotational energy that is due to the leading order
contribution to the electromagnetic energy, the charge contribution
to the total energy, $\frac14 \,{\frac {{Q}^{2}}{M}}$.

\section{Conclusion}

In this paper, it was proposed that the magnetic field shedding
model of FRBs can be understood self-consistently if one includes a
metastable KNBH state after the collapse from a progenitor NS. The
magnetic flux is shed by electric discharge {\bf and not because of
} a no hair requirement. The analysis of the last section resolves
some issues and raises more questions. For NS progenitors with
$P_{NS} > 1$ sec, the pulsar death-line indicates that collapse to a
KNBH should occur without any influence from pair creation in the
magnetosphere. This state can discharge electrically thereby
reproducing the magnetic field shedding in the FRB and GRB models
\citep{han97,fal14,zha14}. The argument has a range of applicability
for $10^{11}\rm{G} < B_{NS} < 10^{13}\rm{G}$.
\par The vast majority of pulsars are likely dead and are generally
of low luminosity and not detected. However, it is believed that
spin up accretion in a binary system can recycle dead pulsars at a
rate $\approx 1.5\, - \,3.0\times 10^{-3}\rm{yr}^{-1}$
\citep{des95}. However, it is known that accretion can also spin
down pulsars \citep{cha97}. This spin-down will increase the mass of
the NS and decrease centrifugal force. For a heavier NS, centrifugal
force might be required to prevent gravitational collapse and spin
down accretion might initiate catastrophic collapse to a BH. The
catastrophic collapse rate of dead pulsars would likely be the same
order of magnitude as the recycle rate of dead pulsars, $\approx
1.5\, - \,3.0\times 10^{-3}\rm{yr}^{-1}$, which agrees with the FRB
birth rate, $ 10^{-3}\, \mathrm{gal}^{-1} \, \mathrm{yr}^{-1}$. The
reducible mass associated with the magnetic field in Equation (17)
would be manifest as an electromagnetic pulse that has been proposed
to power the FRB either from high energy pair plasma in a shock wave
or by coherent emission from particle bunching in the strong
electromagnetic wave \citep{fal14,zha14}. From Figure 1, the
metastable KNBH low spin state has a stored electric energy that far
exceeds the stored magnetic energy. Being longitudinally polarized,
the release of this energy is not directly into electromagnetic
waves. It has been proposed (and extensively modeled) that the
reducible mass in Equation (17) is extracted by a fireball of pair
plasma in models of GRBs with afterglow \citep{ruf01}. The energy of
FRB 150418 and its afterglow is $\sim10^{49}\, -\,10^{50}$ergs,
consistent with the stored electric energy in Figure 1
\citep{kea16,zha16}. It may also be that FRB 150418 represents a
different class of FRB and this is a numerical coincidence
\citep{kea16}.

\par For the supramassive NS invoked in
\citet{fal14,zha14} the death-line argument cannot be used to
validate this scenario since they require $P_{NS} < 0.1$ s in order
to provide adequate centrifugal support. Force-free simulations
indicate that the flux will be shed by prompt reconnection and no
KNBH meatastable states form \citep{leh12}. Although the ``electric
field pruning" that was implemented might render such a simulation
incapable of finding KNBHs and their ``electric" magnetospheres. In
any event, these results raise the question of whether there is any
observational evidence of highly magnetized black hole
magnetospheres associated with stellar mass black holes. There is a
major prediction of such a scenario, the magnetosphere will produce
a tenuous outflow of electron-positron plasma at a highly
relativistic velocity in the form of a jet or wind
\citep{bla77,lyu11}. We find no direct observations of highly
relativistic winds from black holes, yet there are numerous
observations indicating non-relativistic or mildly relativistic jets
and ejections. The two most publicized examples of highly
superluminal motion have not stood the test of time and better data.
The direct observations of component motion in Galactic BHs show
subluminal or slightly superluminal motion, not the highly
superluminal motion expected from black hole driven magnetically
dominated outflows. For example, the discrete ejecta in GRO J1655-40
are no longer considered to be highly superluminal and are now
believed to move $\sim 037$c \citep{fol06}. The parallax determined
distance measurement to GRS~1915+105 is $8.6^{+2.0}_{-1.6}$ kpc
\citep{rei14}. This changes the intrinsic velocity of the discrete
ejections to 0.65c - 0.81c. The parallax measurements and the jet
asymmetry arguments in \citet{rib04} indicate that the compact jet
in GRS~1915+105 propagates with 0.07c $<v< $ 0.37c. The best case
for moderately relativistic ejection in a Galactic BH is a Lorentz
factor of 2 in one of the flares of XTE J1550-564, the other flares
are considerably slower \citep{oro11}. This result, even though it
is not the highly relativistic speed expected from an event horizon
magnetospheric outflow, still depends on a distance estimate that
has not been verified by a parallax observation. One could argue
that the slow outflow velocities are the manifestation of mass
loading from the enveloping medium \citep{mur14}. But,
interferometric observations of ejecta in Galactic BHs agree with
radio light curves - the speed $\approx$constant with no abrupt
luminosity change near the time of ejection \citep{dha00,pun13}.
Baryon loading (if it occurs) seems to occur at the source. Yet, the
putative event horizon magnetosphere is baryon poor. The
observations seem to indicate that the event horizons of Galactic
black holes are either weakly magnetized or non-magnetized.
Apparently, any significant magnetization that they had was a brief
transient or was shed during formation.
\section*{Acknowledgments} We would like to thank Tong Liu for valuable
discussions.

\end{document}